# *In-situ* calibration of forward hadron calorimeters of CMS at LHC

V.M. Biryukov[♦]

Institute for High Energy Physics, 142281 Protvino, Russia

## Abstract

Physical possibility for bending the LHC protons (or ions) a huge angle of 1-20 degrees in the energy range of 0.45 to 7 TeV by means of a bent channeling crystal of Silicon or Germanium is demonstrated. Such an application can be useful for calibration of CMS (or ATLAS) calorimeters *in situ* by the LHC beam of precisely known energy. We show by simulations that such an application would be feasible at the LHC, and report the experience of IHEP Protvino in bending 70 GeV protons by 9 degrees (150 mrad) during 10 years in 1994-2004 experiments.

## Introduction

Calibration of calorimeters manufactured for the CMS and ATLAS experiments of the LHC is essential, however it is difficult and limited in value if done in the conditions different from those foreseen for the actual work in the LHC: at very different energy, in different magnetic field, on a different apparatus etc.

It would be ideal if such a calibration can be done *in situ* with LHC primary beam of precisely known energy in the environment exactly the same as used for the actual work. To realize this, one would need to bend the beam circulating in the LHC by a huge angle, order of 1-10 degrees (17-170 mrad) [1]. Beam intensity as low as 1-1000 p/s would be sufficient for calibration purposes.

Such a bending is possible indeed, but only by extreme, ~1000-Tesla-strong fields of bent crystal. The technique of particle beam channeling by bent crystals is well established at accelerators [2]. Broad experience is obtained with bent channeling crystals

---

[♦] http://mail.ihep.ru/~biryukov/

at IHEP, CERN SPS, Tevatron, and RHIC over recent decades [2-15]. For instance, bent crystals are largely used for extraction of 70-GeV protons at IHEP (Protvino) [11-15] with efficiency reaching 85% at intensity of $10^{12}$ particle per second, steered by silicon crystal of just 2 mm in length [12]. Much of the IHEP physics program relies on crystal channeled beams regularly used since 1989. Following the successful experiments on crystal channeling in accelerator rings, there has been a strong interest to apply channeling technique in a TeV-range collider for beam extraction or collimation [2-8,16-18].

## IHEP experiment on crystal bending of 70 GeV protons by 150 mrad

The possibility of beam bending by a channeling crystal through very large angles, order of 100 mrad, was studied at Protvino. In 1994-2004 IHEP has operated a channeling crystal 100 mm long to bend 70 GeV beam a huge angle of 150 mrad (9 degrees!) [19,20]. This was done with the purpose to organize over a short base a non-traditional beam line. A 100 mm long Si (110) crystal deflected $10^6$ proton/s beyond the 2-meter iron-concrete shield over 20 m base, out of about $10^{10}$ proton/s hitting the crystal. The crystal efficiency was in agreement with calculation. The orientation-independent component of the signal (background particles) did not exceed 3% of the channeled beam. The scheme of 150-mrad crystal beam line operated at IHEP at $10^6$ proton/s in 1994-2004 is shown in Figure 1.

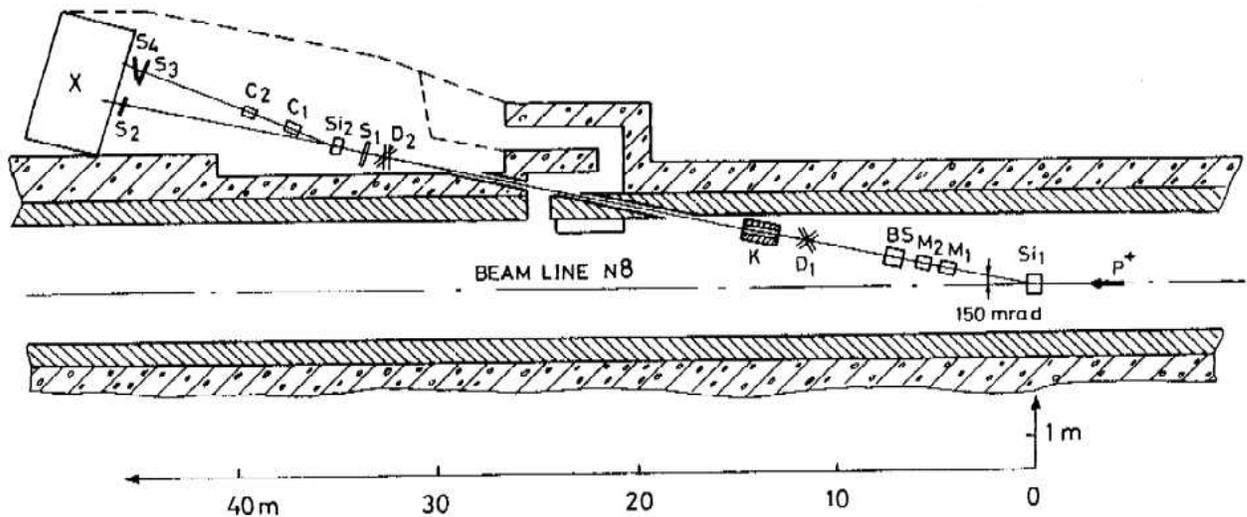

Fig. 1. The scheme of crystal beam line and experimental setup: $Si_1, Si_2$ – deflecting and testing crystals, $M_1, M_2$ – corrector magnets, BS – beam stopperp, $D_1, D_2$ – proportional chambers, K – a collimator, $S_1 - S_4$ – scintillator countes, C1, C2 – microstrip detector stations, X – a beam absorber.

This crystal beam line was used mostly for our studies of channeling. It consumed practically no power and allowed one to work in parallel with other beam lines without affecting other physical set-ups operation.

Besides this 150-mrad bent crystal used in 1994-2004, another example in IHEP experience was an 85-mrad bent crystal used for extraction in 1989-1999 over 10 years until a new crystal replaced it [2].

## Simulations for the LHC beam bending by 1-20 degrees

In order to find out whether one can extrapolate the 70 GeV experience onto the LHC case of much higher energy range, we performed Monte Carlo simulations applying the code [21-23] used in many other channeling experiments [6-15]. Details of the theoretical calculations can be found, e.g., in book [2].

Our preliminary study assumed a crystal of 50 cm size, which is a reasonable figure available on the market in Russia and Europe. Earlier, IHEP has used up to 15 cm long Si crystals for channeling.

An example of the calculation of efficient transmission of LHC beam with Silicon crystal at the energies of 450 GeV (injection level in the LHC) and 1000 GeV is shown in Figure 2. Respectively, bending angles up to about 7 and 11 degrees can be realized with Si crystal at these energies.

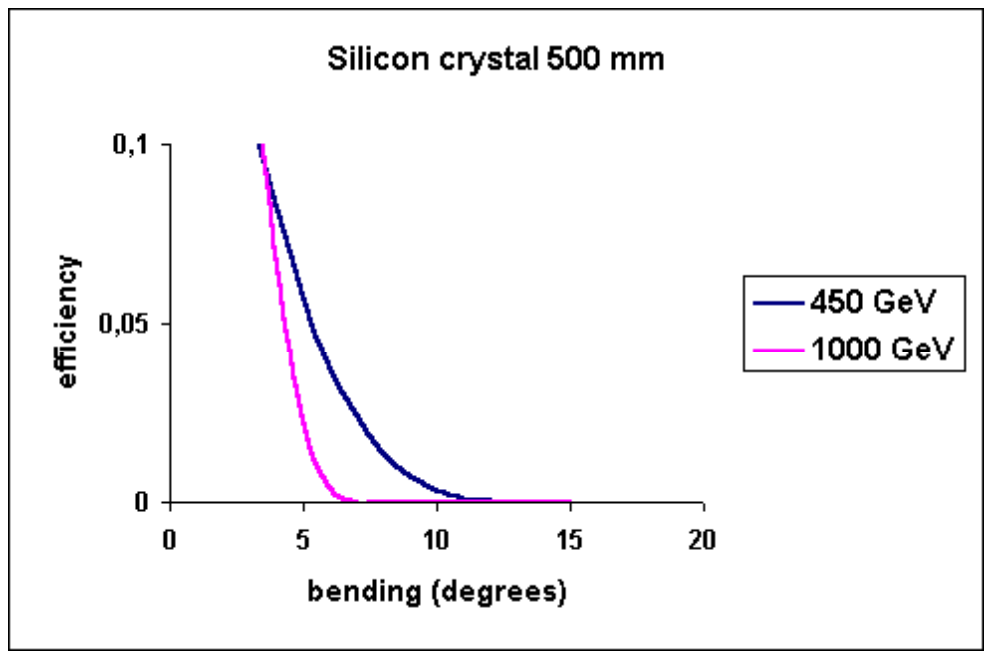

**Figure 2 Efficiency of the LHC proton beam bending with a single *Si* crystal 500 mm long as a function of the bending angle (degrees).**

We have run computations for Silicon and Germanium crystals of 50 cm size and the LHC energy range from injection, 450 GeV, through the top energy, 7 TeV. Some preliminary results are shown on Figure 3 for the bending angle achievable with a single Si crystal of 50 cm length as a function of the LHC beam energy.

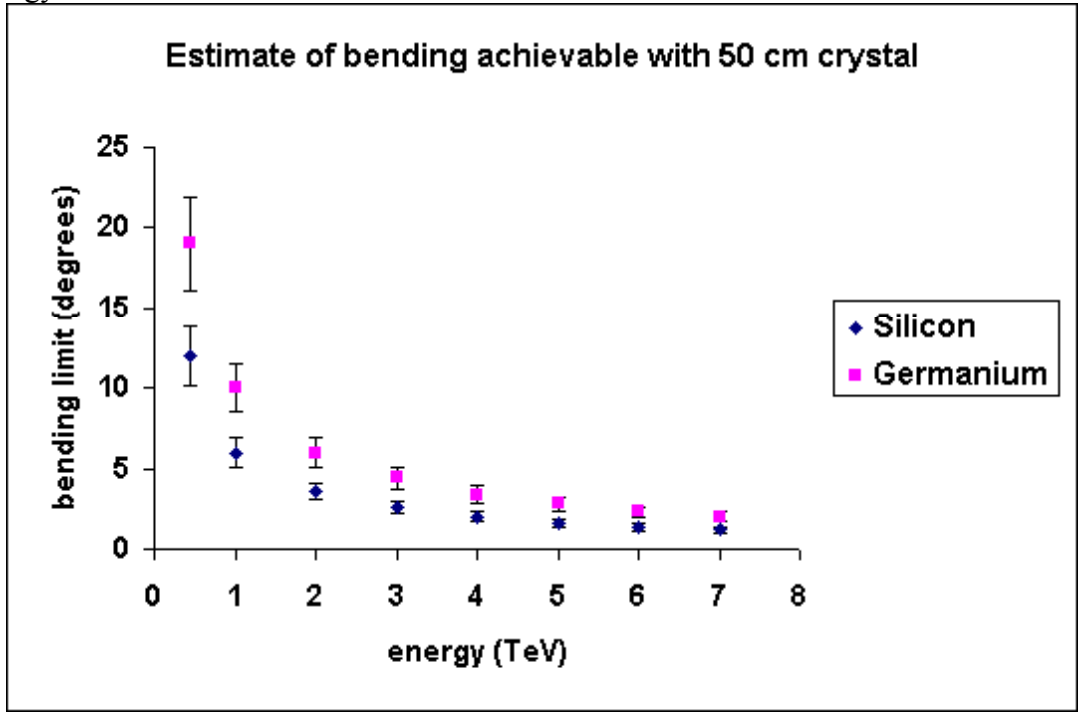

**Figure 3 Bending angle achievable with a single Si crystal of 50 cm length as a function of the LHC beam energy.**

This opens the possibility for calibration of CMS (or ATLAS) HF and HE calorimeters *in situ* by channeling the LHC protons of full energy range of 0.45 to 7 TeV right into calorimeters, irradiating them cell by cell, with a crystal bent a very large angle, ~10 to 350 mrad, or 1 to 20 degrees.

In principle, the deflection angle of crystal can be varied *in situ*. Several approaches were exploited at IHEP and CERN, for instance by N. Doble et al. at the SPS [2,].

The same technique can be used with heavy ions if they fill the LHC ring, or even with positrons provided there would be a positron source filling into the LHC ring!

It is essential to notice that at the energies lower than 450 GeV, with the crystal of reasonable length like 50 cm one can realize bending angles essentially bigger than even 20 degrees! In this case the LHC ring would have to be operated at a magnetic field lower than in the injection regime provided that power supplies of magnets tolerate it.

**Conclusion**

Crystal channeling is a well established technique which has been exploited at IHEP for decades. In particular, IHEP bent 70 GeV protons an angle of 9 degrees (150 mrad) at the intensity of $10^6$ proton/s over 10 years since 1994.

The CMS and ATLAS calorimeters can be calibrated *in situ* by channeling the LHC primary protons of precisely known energy right into calorimeters, irradiating them cell by cell with a crystal bent a very large angle, ~1 to 20 degrees and beyond, exploiting the full energy range of 0.45 to 7 TeV. The same can be done with heavy ions, or even with positrons provided the LHC ring stores these particles.

**Acknowledgement**. This work was supported by INTAS-CERN grant 03-52-6155.

## References

[1] V.I. Kryshkin, N. Akchurin, private communication (2004).
[2] V.M. Biryukov, Yu.A. Chesnokov and V.I. Kotov, *"Crystal Channeling and its Application at High Energy Accelerators"* (Springer, Berlin, 1997)
[3] H. Akbari et al. *Phys. Lett. B* 313 (1993) 491
[4] X. Altuna et al., *Phys. Lett. B* **357**, 671 (1995).
[5] A. Baurichter, et al. *Nucl.Instrum.Meth.B* **164-165**: (2000) 27-43
[6] R.A. Carrigan et al., *Nucl. Instrum. Meth.* B **119** (1996) 231
[7] R.A.Carrigan, Jr., et al. *Phys. Rev. ST AB* **1**, 022801 (1998)
[8] R.A. Carrigan et al. *Phys. Rev. ST Accel. Beams* **5** (2002) 043501.
[9] R.P. Fliller III, A. Drees, D. Gassner, L. Hammons, G. McIntyre, S. Peggs, D. Trbojevic, V. Biryukov, Y. Chesnokov, V. Terekhov, *AIP Conf. Proc.* **693** (2003) 192-195
[10] R.P. Fliller III, A. Drees, D. Gassner, L. Hammons, G. McIntyre, S. Peggs, D. Trbojevic, V. Biryukov, Y. Chesnokov, V. Terekhov, *Nucl. Instr. and Meth. B*, in press
[11] A.G. Afonin et al. *Phys. Lett. B* **435** (1998) 240-244.
[12] A.G. Afonin et al. *Phys. Rev. Lett*. **87**, 094802 (2001)
[13] V.M. Biryukov et al. Rev. Sci. Instrum. **73** (2002) 3170-3173
[14] A.G. Afonin et al. *JETP Lett.* **74** (2001) 55-58
[15] A.G. Afonin et al. *Instrum. Exp. Tech.* **45**(4) (2002) 476
[16] V. Biryukov. *Phys. Rev. Lett.* **74** (1995) 2471.
[17] V.M. Biryukov, V.N. Chepegin, Yu.A. Chesnokov, V. Guidi, W. Scandale, *Nucl. Instr. and Meth. B*, in press
[18] V.M. Biryukov, A.I. Drozhdin, N.V. Mokhov. PAC 1999 Proc. (New York), p.1234. FERMILAB-Conf-99-072 (1999), "On Possible Use of Bent Crystal to Improve Tevatron Beam Scraping".
[19] V.M. Biryukov et al. IHEP Preprint 95-14 (1995).
[20] V.M. Biryukov et al. PAC Proceedings (Dallas, 1995).V.Biryukov. CERN SL Note 93-74 AP (1993). "Crystal Channeling Simulation. CATCH 1.4 User's Guide".
[22] V. Biryukov. *Phys. Rev. E* **51** (1995) 3522.
[23] V. Biryukov. *Phys. Rev. Lett*. **74** (1995) 2471.
[24] N. Doble, P. Grafstrom, L. Gatignon. *Nucl. Instr. and Meth. B* **119** (1996) 181